\documentclass[a4paper, amsfonts, amssymb, amsmath, reprint, showkeys, nofootinbib, twoside]{revtex4-1}
\usepackage[english]{babel}
\usepackage[utf8]{inputenc}
\usepackage[colorinlistoftodos, color=green!40, prependcaption]{todonotes}
\usepackage{amsthm}
\usepackage{ulem}
\usepackage{multirow} 
\usepackage{xcolor}
\usepackage{graphicx}
\usepackage{caption}
\usepackage[T1]{fontenc}
\usepackage[pdftex, pdftitle={Article}, pdfauthor={Author}]{hyperref} % For hyperlinks in the PDF

\definecolor{purple}{rgb}{0.58,0.0,0.83}

\definecolor{blue(pigment)}{rgb}{0.2, 0.2, 0.6}

\usepackage{scalerel}
\usepackage{tikz}
\usetikzlibrary{svg.path}

\definecolor{orcidlogocol}{HTML}{A6CE39}
\tikzset{
  orcidlogo/.pic={
    \fill[orcidlogocol] svg{M256,128c0,70.7-57.3,128-128,128C57.3,256,0,198.7,0,128C0,57.3,57.3,0,128,0C198.7,0,256,57.3,256,128z};
    \fill[white] svg{M86.3,186.2H70.9V79.1h15.4v48.4V186.2z}
                 svg{M108.9,79.1h41.6c39.6,0,57,28.3,57,53.6c0,27.5-21.5,53.6-56.8,53.6h-41.8V79.1z M124.3,172.4h24.5c34.9,0,42.9-26.5,42.9-39.7c0-21.5-13.7-39.7-43.7-39.7h-23.7V172.4z}
                 svg{M88.7,56.8c0,5.5-4.5,10.1-10.1,10.1c-5.6,0-10.1-4.6-10.1-10.1c0-5.6,4.5-10.1,10.1-10.1C84.2,46.7,88.7,51.3,88.7,56.8z};
  }
}

\newcommand\orcidicon[1]{\href{https://orcid.org/#1}{\mbox{\scalerel*{
\begin{tikzpicture}[yscale=-1,transform shape]
\pic{orcidlogo};
\end{tikzpicture}
}{|}}}}

\begin{document}

\title{Revealing some cosmological aspects of Kaniadakis entropy}

\author{Miguel Cruz$^1$\orcidicon{0000-0003-3826-1321}}
\email{miguelcruz02@uv.mx}

\author{Samuel Lepe$^{2}$\orcidicon{0000-0002-3464-8337}}
\email{samuel.lepe@pucv.cl}

\author{Joel Saavedra$^2$\orcidicon{0000-0002-1430-3008}}
\email{joel.saavedra@pucv.cl}

\affiliation{$^1$Facultad de F\'{\i}sica, Universidad Veracruzana 91097, Xalapa, Veracruz, M\'exico,\\
$^2$Instituto de F\'\i sica, Pontificia Universidad Cat\'olica de Valpara\'\i so, Casilla 4950, Valpara\'\i so, Chile.}

\date{\today}

\begin{abstract}
Adopting the modifications induced by the truncated version of the Kaniadakis entropy on the Friedmann equations, we explore some relevant aspects of this cosmological scenario at the background level. We analyze the constraint imposed on the parameter $K$ obtained from the accelerated cosmic expansion condition, and we also study the role of such a parameter as a cosmological constant. 
\end{abstract}

%\pacs{98.80.Cq}

\maketitle

%%%%%%%%%%%%%%%%%%%%%%%%%%%%%%%%%%
\section{Introduction} 
%%%%%%%%%%%%%%%%%%%%%%%%%%%%%%%%%%%
The study of late-time cosmic acceleration remains one of the most intriguing challenges in modern cosmology. Although the $\Lambda$CDM model has been remarkably successful in describing observational data \cite{planck1518}, fundamental questions about the nature of dark energy and cosmic expansion persist. In order to unveil the unknown nature of the dark sector, the intersection of gravity and thermodynamics has been a fertile ground to explore.

The starting point of this line of reasoning is given by Bekenstein and Hawking's groundbreaking work on black hole thermodynamics \cite{bekenhaw}, where the entropy depends on non-additive statistics since it is written in terms of the area of the event horizon and not the volume of the system under consideration, as usual. The gravity-thermodynamic conjecture, which suggests a deep connection between gravitational dynamics and thermodynamic principles, has provided valuable insights into the nature of spacetime and cosmic evolution; under the assumption of the proportionality of entropy and horizon area, in Ref. \cite{jacobson} it was demonstrated that the Einstein equation is indeed an equation of state since it can be obtained from the relation $\delta Q = TdS$, which is a fundamental relation connecting heat, entropy, and temperature. This is the well-known Clausius relation of classical thermodynamics \cite{callen}.

As a case of interest, the analogy for dynamic cosmological models of the formalism mentioned above was carried out accordingly. The Friedmann equations for a homogeneous and isotropic universe were derived by applying the first law of thermodynamics to the apparent horizon; such derivation was based on the already known entropy-area linear relationship \cite{hayward, cai}. This subject of investigation is actually known as {\it entropic cosmology}, and we refer the reader to the reference \cite{entropic}. This framework implies interpreting the apparent horizon as a thermodynamic system, requiring the existence of a fundamental equation from which all the thermodynamic properties of the horizon can be obtained \cite{quevedo}. See also references given in \cite{creation}, where the thermodynamics and cosmology associated with the apparent horizon are studied incorporating matter creation effects. An interesting reference with a pedagogical perspective in the treatment of the apparent horizon can be found in \cite{faraoni}.

On the other hand, if we proceed in the opposite direction, the gravity-thermodynamic conjecture also admits the investigation of the first law of thermodynamics and entropy of the apparent horizon in the presence of cosmological fluids in diverse theories of modified gravity\footnote{Under this perspective the matter fields inside the apparent horizon obey the thermodynamics of an open system since an inward or outward matter flux through the apparent horizon can be considered \cite{flux}.}; see for instance Ref. \cite{sebastiani}, where distinct generalizations for the entropy of the apparent horizon emerged from diverse modified Friedmann equations\footnote{A unifying feature of the modified gravity models studied is the preservation of second-order differential equations of motion.}. However, an interesting fact is that the modifications obtained for the entropy of the apparent horizon are given in terms of the Bekenstein-Hawking entropy, $S_{\mathrm{BH}}$, i.e., $S_{\mathrm{A}} = f(S_{\mathrm{BH}})$ with $S_{\mathrm{BH}}\propto A$, being $f$ an arbitrary function determined by the gravity theory under consideration. The exploration of generalized entropy formulations \cite{entropies} has opened new avenues to understand the dynamics of cosmic evolution, with the Kaniadakis entropy emerging as a particularly intriguing framework \cite{kaniadakis}.\\ 

The Kaniadakis entropy represents a significant generalization of standard Boltzmann-Gibbs statistics. It is constructed to be consistent with the fundamental symmetries of special relativity and preserves crucial mathematical and physical properties of thermodynamic systems. Consequently, this framework has been successfully applied to describe a wide range of physical phenomena exhibiting power-law behavior \cite{silva}. When applied to cosmological scenarios, particularly to describe the universe's late-time evolution, it yields profound implications. By incorporating Kaniadakis entropy into the holographic principle, one can derive modified Friedmann equations. These modifications can generate an effective dark energy component endogenously, driven by the generalized entropy of the spacetime itself. As a result, this framework provides a viable explanation for cosmic acceleration and has the potential to alleviate major observational tensions, such as the Hubble constant discrepancy, without the need to postulate exotic and physically unverified forms of matter or energy \cite{saridakis, salehi1}.\\

This letter focuses on the late-time cosmological implications of the Kaniadakis entropy formalism, and we examine the consequences of the modifications to the standard cosmological equations introduced by this entropy. Of particular interest is how this framework naturally incorporates features that could explain the observed acceleration without explicitly introducing a cosmological constant. The structure of the letter is as follows: In the next section, the generalities of the cosmological equations at the background level are described; the conditions to have accelerated cosmic expansion in the model are also studied. The section \ref{sec:kaniadakis} is devoted to exploring some possible physical interpretations of the Kaniadakis parameter and its role as a cosmological constant; we compare numerically our results to some values obtained in the literature by some collaborations. Finally, in section \ref{sec:final}, we give some final comments on our work. 

%%%%%%%%%%%%%%%%%%%%%%%%%%%%%%%%%%%%%%%%%%%%%%%%%%%%%%%%%%%
\section{Cosmological model}
\label{sec:cm}
%%%%%%%%%%%%%%%%%%%%%%%%%%%%%%%%%%%%%%%%%%%%%%%%%%%%%%%%%%
In this section, we provide some highlights of the cosmological model to consider in our analysis for a Friedmann-Lema\^itre-Robertson-Walker (FLRW) space-time with null spatial curvature in the context of the apparent horizon description, $8\pi G = k_{\mathrm{B}}=c = 1$ units will be used. The subscript zero will denote the evaluation of cosmological quantities at the present time. The main advantage of the Kaniadakis entropy is its dependence on a single parameter, $K$, which quantifies deviations from standard statistical mechanics. Its definition is given as \cite{abreu}
\begin{equation}
    S_{\mathrm{K}}=-\sum^{W}_{i=1}\frac{P^{1+K}_{i}-P^{1-K}_{i}}{2K}, \label{eq:def}
\end{equation}
where $P_{i}$ represents the probability system to occupy a specific microstate and $W$ is the total number of configurations. Within the context of black hole physics, we assume $P_{i} = 1/W$ \cite{pavon}, using the fact that Boltzmann-Gibbs entropy is $S=\ln (W)$, it yields $W=\exp (S_{\mathrm{BH}})$, therefore we obtain from Eq. (\ref{eq:def})
\begin{equation}
    S_{\mathrm{K}}=\frac{1}{K}\sinh (KS_{\mathrm{BH}}), \label{eq:kaniadakis}
\end{equation}
where $K$ is the Kaniadakis parameter restricted to the interval, $0 < K < 1$ and $S_{\mathrm{K}\rightarrow 0}=S_{\mathrm{BH}}$, in such limit the Bekenstein-Hawking entropy is recovered. As mentioned above, $S_{\mathrm{BH}}=A/4$, being $A$ the area of the apparent horizon, $A=4\pi R^{2}_{A}$ with the radius given by $R_{A}=H^{-1}$, as usual for a flat spacetime, the radius of the apparent horizon becomes the Hubble radius. It is worthy to mention that in Ref. \cite{moradpour} it was shown that Kaniadakis entropy can be related to other generalized entropies as the Tsallis entropy, $S^{\mathrm{T}}_{\mathrm{Q}}$. A notable feature common to both Tsallis and Kaniadakis entropy is that each is dependent on a single defining parameter. For the Tsallis entropy  the free parameter is dubbed as $Q$ and codifies the non-extensivity properties of the entropy\footnote{The Tsallis entropy is given as 
\begin{equation*}
  S^{\mathrm{T}}_{\mathrm{Q}} =\frac{2}{1-Q}\exp \left[\frac{(1-Q)}{2}S_{\mathrm{BH}}\right]\sinh \left[\frac{(1-Q)}{2}S_{\mathrm{BH}}\right].   
\end{equation*}}, this non-standard description is broadly applicable to complex systems with long-range correlations or non-ergodic properties. In this case the Eq. (\ref{eq:def}) is found to be 
\begin{equation}
    S_{\mathrm{K}}=\frac{S^{\mathrm{T}}_{\mathrm{1+K}}-S^{\mathrm{T}}_{\mathrm{1-K}}}{2}, \label{eq:def2}
\end{equation}
i.e., there exists a correspondence between the parameters $Q$ and $K$. Tsallis entropy also shares connections with other well-known entropies, including the R\'enyi and Sharma-Mittal formulations, therefore these scenarios do not provide new information w.r.t. Kaniadakis entropy. Additionally, the alternative formulation of the Tsallis entropy, $S_{T} = \gamma A^{\delta}$ (which depends on two free parameters), does not satisfy the equation (\ref{eq:def2}) and this also happens for the Barrow entropy defined as, $S_{B} = A^{1+\Delta}$ \cite{depranou}. In both cases the parameters $\delta$ and $\Delta$ measure quantum-gravitational effects that impose the fractal structure on the surface of the black hole. However, a better understanding of the parameters $\gamma$, $\delta$ and $\Delta$ could be provided only in the context of quantum-gravitational scenarios, a situation that still seems distant. We would like to emphasize that while Barrow entropy has direct links to quantum gravity phenomenology, Tsallis entropy is not confined to a quantum gravitational context and provides a more general framework for systems that do not satisfy the additivity axiom of standard statistical mechanics. \\

If one expects the generalized entropy (\ref{eq:kaniadakis}) to be close to the standard case, then we must have $K\ll 1$ to achieve such condition, as mentioned before $0 < K < 1$. Small values of the parameter $K$ are also consistent with the holographic dark energy scenario as found in Ref. \cite{depranou}; then Eq. (\ref{eq:kaniadakis}) takes the following form after a series expansion
\begin{equation}
    S_{\mathrm{K}} = S_{\mathrm{BH}}+\frac{K^{2}}{6}S^{3}_{\mathrm{BH}}+\mathcal{O}(K^{4}),\label{eq:kanapprox}
\end{equation}
as can be seen, the zeroth-order term is the usual entropy and the second term is the lowest-order Kaniadakis correction to the entropy, then it is clear to observe that the limit, $K\rightarrow 0$, recovers the BH entropy. See Appendix (\ref{sec:valuek}) for a brief discussion regarding the validity of Eq. (\ref{eq:kanapprox}). From the gravity-thermodynamics conjecture, it was found that entropy (\ref{eq:kaniadakis}) leads to the following set of cosmological equations modulo an integration constant, usually denoted as $\Lambda$, for a flat spacetime \cite{saridakis}
\begin{align}
    & 3H^{2}\left[\cosh\left(KS_{\mathrm{BH}}\right)-KS_{\mathrm{BH}}\mbox{shi}\left(KS_{\mathrm{BH}}\right) \right] = \rho_{\mathrm{m}},\label{eq:fried2}\\
    & \dot{H}\cosh\left(KS_{\mathrm{BH}}\right) = -\frac{1}{2}(\rho_{\mathrm{m}} + p_{\mathrm{m}})\label{eq:accel2},
\end{align}
where $\mbox{shi}(x)$ is a mathematical odd function of $x$ whose definition is $\mbox{shi}(x)=\int^{x}_{0}\frac{\sinh(x')}{x'}dx'$ and the dot represents derivatives with respect to cosmic time. It is worth mentioning that under the series expansions of the function $\cosh{(x)} = 1+\frac{x^{2}}{2}+\frac{x^{4}}{24}+...$ and $\mbox{shi}(x)=x+\frac{x^{3}}{18}+\frac{x^{5}}{600}+...$, the dynamical equations (\ref{eq:fried2}) and (\ref{eq:accel2}) coincide with those obtained in Ref. \cite{sheykhi}, as discussed in Ref. \cite{salehi}.

In Ref. \cite{saridakis} the role of the integration constant $\Lambda$ was discussed; the case $\Lambda=0$ imposes the condition $K\neq 0$ since $K=0$ only describes a CDM scenario. For this scenario, $K$ should vary consistently with the observational range of $\Omega_{\mathrm{m,0}}$. However, in such a study, the case $\Lambda \neq 0$ was chosen since it reproduces the thermal history of the universe (matter and dark energy epochs), and under the limit $K\rightarrow 0$, the $\Lambda$CDM model is recovered. Additionally, $\Lambda=0$ leads to a dark energy sector that behaves as phantom dark energy in the past, and only in the far future, a de Sitter phase dominates the cosmic evolution; as stated by the authors of \cite{saridakis}, this latter scenario is less attractive. During the study of Ref. \cite{saridakis}, the presence of $\Lambda$ and the parameter $K$ was studied separately. We will discuss this aspect of the model below. An interesting fact of the set of equations (\ref{eq:fried2})-(\ref{eq:accel2}) is that the geometry of the field equations is modified due to the generalization of the entropy; then these emergent geometric terms can be associated to a dark energy sector whose origin is known; for $K=0$, we recover the CDM scenario and thus we must impose the condition $K\neq 0$ in order to have extra contributions.
If we assume a barotropic EoS for the matter sector, $p_{\mathrm{m}}=\omega \rho_{\mathrm{m}}$, together with Eq. $(\ref{eq:accel2})$, we can write the deceleration parameter $q(t)=-1-\frac{\dot{H}}{H^{2}}$, as follows. 
\begin{widetext}
\begin{equation}
    1+q(t) = \frac{1}{2}(1+\omega)\frac{\rho_{\mathrm{m}}}{H^{2}\cosh\left(KS_{\mathrm{BH}}\right)}=\frac{3}{2}(1+\omega)\left[1-\frac{KS_{\mathrm{BH}}\mbox{shi}\left(KS_{\mathrm{BH}}\right)}{\cosh\left(KS_{\mathrm{BH}}\right)}\right], \label{eq:decel}
\end{equation}
\end{widetext}
where Eq. (\ref{eq:fried2}) was also considered. Taking the derivative w.r.t. cosmic time of Eq. (\ref{eq:fried2}) we get $\dot{\rho}_{\mathrm{m}} = -6H^{3}(1+q)\cosh\left(KS_{\mathrm{BH}}\right)$, then if we insert the term given in the first equality of Eq. (\ref{eq:decel}), we obtain the usual conservation equation for the matter sector, $\dot{\rho}_{\mathrm{m}}+3H(1+\omega)\rho_{\mathrm{m}}=0$. 

In order to have an accelerated cosmic expansion, the condition $q(t) < 0$ must be satisfied. According to the expression (\ref{eq:decel}) for $K=0$, we obtain $q=(1/2)(1+3\omega)$, which coincides with the standard cosmology for a single fluid described by the parameter state $\omega$. The values $K=0$ and $\omega=0$ in the equation (\ref{eq:decel}) recover the CDM case, $q=1/2$. Some comments are in order. Evaluating Eq. (\ref{eq:decel}) at present time with $\omega=0$, yields
\begin{equation}
    q_{0} = \frac{1}{2}\left(1-3\left[\frac{\mathcal{K}\mbox{shi}\left(\mathcal{K}\right)}{\cosh\left(\mathcal{K}\right)}\right]\right), \label{eq:decel1}
\end{equation}
being $\mathcal{K}$ a constant defined as follows, $\mathcal{K}:=K\pi H^{-2}_{0}$. As can be seen, the deceleration parameter at present time can be estimated with the use of the series expansions $\cosh{(\mathcal{K})} = 1+\frac{\mathcal{K}^{2}}{2}+\frac{\mathcal{K}^{4}}{24}+...$ and $\mbox{shi}(\mathcal{K}) = \mathcal{K}+\frac{\mathcal{K}^{3}}{18}+\frac{\mathcal{K}^{5}}{600}+...$, once the values for the Hubble constant, $H_{0}$, and the Kaniadakis parameter, $K$, are given. This represents a great simplification since obtaining cosmological constraints for $q$ from astrophysical observations implies establishing physical observables such as luminosity distance which is given as $d_{L}(z) = (1+z)\int^{z}_{0}dz'/H(z')$; additionally, $H$ is related to $q$ by means of $H(z)=H_{0}\exp[\int^{z}_0(1+q(u))d\ln (1+u)]$. The aforementioned set of equations is in general solvable only under the consideration of parameterizations for $q$. From now on, $z$ is the redshift given by the usual formula $1+z=a(t_{0})/a(t)$, being $a(t_{0})$ the value of the scale factor at present time, $t_{0}$.

\begin{figure}[htbp!]
    \includegraphics[width=0.5\textwidth]{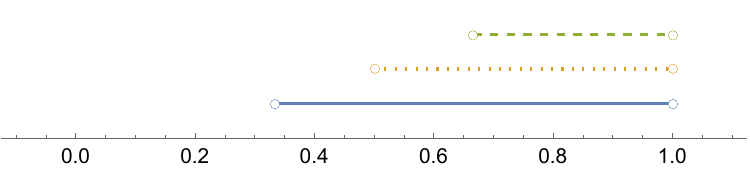}
\caption{Validity region for the term $\frac{KS_{\mathrm{BH}}\mbox{shi}\left(KS_{\mathrm{BH}}\right)}{\cosh\left(KS_{\mathrm{BH}}\right)}$ for which $q(t)<0$ with $\omega=0$ (solid line), $\omega=1/3$ (dotted line) and $\omega=1$ (dashed line).}
\label{fig:decel}
\end{figure}

Taking into account the condition $q(t) < 0$, the values taken by the term $\frac{KS_{\mathrm{BH}}\mbox{shi}\left(KS_{\mathrm{BH}}\right)}{\cosh\left(KS_{\mathrm{BH}}\right)}$ must be restricted. In Fig. (\ref{fig:decel}), we show the intervals of $\frac{KS_{\mathrm{BH}}\mbox{shi}\left(KS_{\mathrm{BH}}\right)}{\cosh\left(KS_{\mathrm{BH}}\right)}$ leading to $q(t) < 0$ in Eq. (\ref{eq:decel}); we observe that for a matter content given by radiation ($\omega=1/3$) or stiff matter ($\omega=1$) the possible values of $\frac{KS_{\mathrm{BH}}\mbox{shi}\left(KS_{\mathrm{BH}}\right)}{\cosh\left(KS_{\mathrm{BH}}\right)}$ lie within the interval obtained for CDM ($\omega=0$). Therefore, the range of the function $\frac{KS_{\mathrm{BH}}\mbox{shi}\left(KS_{\mathrm{BH}}\right)}{\cosh\left(KS_{\mathrm{BH}}\right)}$ must be constrained to the half-open interval $(1/3,1/2]$ to have a physically consistent cosmic evolution, i.e., we must exclude radiation and stiff matter as possible sources of the accelerated cosmic expansion, as is well known. This latter condition is very restrictive for the $K$ parameter. As expected, the value of the parameter state $\omega$ that describes the content of the universe determines the type of cosmic evolution. As observed, accelerated cosmic evolution can be obtained from Kaniadakis cosmology for a universe filled with a CDM fluid without needing a dark-energy contribution.  

If we consider the expression (\ref{eq:kanapprox}) for the Kaniadakis entropy, we can compute its derivative w.r.t. time, yielding 
\begin{equation}
   \dot{S}_{\mathrm{K}} =\frac{2\pi}{H}[1+q(t)]\left(1+\frac{K^{2}}{2}S_{\mathrm{BH}}\right), \label{eq:positive}
\end{equation}
therefore the second law of thermodynamics \cite{callen}, $\dot{S}_{\mathrm{K}} \geq 0 $, is guaranteed always that $q(t)\geq -1$, i.e., quintessence scenario. This can be fulfilled since for cosmic expansion we have $H(t) > 0$. As commented above, the condition $q(t)\geq -1$ is contained in the expression given in Eq. (\ref{eq:decel}). 

For thermodynamic consistency, two conditions must be satisfied simultaneously by the entropy, positive production, $\dot{S}_{\mathrm{K}} > 0$, and the convexity condition, $\ddot{S}_{\mathrm{K}}<0$. From our previous result, we obtain that $\ddot{S}_{\mathrm{K}}<0$ is satisfied for
\begin{equation}
    \dot{q}(t)<-[1+q(t)]^{2}\left(2\pi+3K^{2}S^{2}_{\mathrm{BH}}\right)\left(\frac{2\pi}{H}+K^{2}S^{2}_{\mathrm{BH}}\right)^{-1},
\end{equation}
No change of sign is expected in the expression given above; then, both conditions can be satisfied by the Kaniadakis entropy as long as $q(t)\geq -1$.

%%%%%%%%%%%%%%%%%%%%%%%%%%%%%%%%%%%%%%%%%%%%%%%%%%%%%%%%%%%%%%%
\section{The Kaniadakis Parameter}
\label{sec:kaniadakis}
%%%%%%%%%%%%%%%%%%%%%%%%%%%%%%%%%%%%%%%%%%%%%%%%%%%%%%%%%%%%%%%

%%%%%%%%%%%%%%%%%%%%%%%%%%%%%%%%%%%%%%%%%%%%%%%%%%%%%%%%%%%%%%
\subsection{Theoretical Interpretations}
%%%%%%%%%%%%%%%%%%%%%%%%%%%%%%%%%%%%%%%%%%%%%%%%%%%%%%%%%%%
The Kaniadakis parameter $K$, introduced as a deformation of the standard Boltzmann-Gibbs entropy, plays an important role in the cosmological model under review by generating geometric corrections capable of driving late-time acceleration. While in our analysis $K$ has been treated as a phenomenological constant, it is natural to inquire whether it could originate from deeper principles based in fundamental theories of spacetime and gravity. One potential address arises from the holographic principle, which postulates that the information content of a gravitational system is encoded on its boundary. In this context, the Kaniadakis entropy may be viewed as an effective description of a microscopic quantum-gravitational regime where standard extensivity and locality break down. Given that the Bekenstein-Hawking entropy, $S_{\mathrm{BH}}$, is already a manifestation of holographic scaling, its deformation via $K$ could reflect corrections to the entropy-area relation induced by a yet-to-be-formulated holographic microscopic theory \cite{Susskind:1994vu}. On the other hand, works in loop quantum gravity and string theory suggest that quantum gravitational degrees of freedom may lead to sub-extensive or super-extensive entropy corrections \cite{Carlip:2014pma}.

Moreover, from a statistical perspective, Kaniadakis entropy can be derived from considerations of anomalous diffusion, scale-invariant phase spaces, or quantum group deformations \cite{Tsallis:1987eu}. These structures appear in the study of quantum black holes, non-commutative geometry, and generalized uncertainty principles—all of which are central in quantum gravity \cite{Kempf:1994su, Maggiore:1993rv}.

In certain non-equilibrium approaches to gravity, deformation parameters like $K$ might emerge as effective couplings that quantify deviations from equilibrium or classical geometry \cite{jacobson}. In this view, $K$ could parameterize the influence of entanglement entropy or quantum coherence on the evolution of the apparent horizon.

These connections, while speculative, suggest that the parameter $K$ may not be merely a phenomenological input but a footprint of more fundamental physics \cite{Housset:2023jcm}. Investigating the emergence of Kaniadakis-type statistics from first principles—whether through holographic dualities, quantum gravity models, or information-theoretic frameworks—remains an exciting direction for future research.

%%%%%%%%%%%%%%%%%%%%%%%%%%%%%%%%%%%%%%%%%%%%%%%%%%%%%%%%%%%
\subsection{Interpreting the role of the $K$-parameter}
\label{sec:kaniadakis2}
%%%%%%%%%%%%%%%%%%%%%%%%%%%%%%%%%%%%%%%%%%%%%%%%%%%%%%%%%%
According to our definitions, the parameter $K$ and the Hubble constant are related by the following equation
\begin{equation}
    KS_{\mathrm{BH},0} = \frac{\pi K}{H^{2}_{0}}.  \label{eq:entropy}
\end{equation}
Consequently, the fractional energy density, $\Omega_{\mathcal{K},0}$, associated with the Kaniadakis parameter will be simply the equation given above divided by 3. Therefore, at second order in $\Omega_{\mathcal{K},0}$, the deceleration parameter (\ref{eq:decel}) with $\omega=0$, takes the form
\begin{equation}
    q_{0} = \frac{1}{2}\left(1-\frac{6\Omega^{3}_{\mathcal{K},0}}{2+\Omega^{2}_{\mathcal{K},0}} \right),
\end{equation}
where $6\Omega^{3}_{\mathcal{K},0}>2+\Omega^{2}_{\mathcal{K},0}$ in order to have $q_{0}<0$. This condition can be fulfilled since $0<\Omega_{\mathcal{K},0}<1$. For $K=0$, the expression given above is consistent with the CDM scenario $q_{0}=1/2$. As commented previously, the accelerated phase for the universe with a CDM sector can be possible due to the presence of the parameter $K$. Using the series expansion mentioned before, we obtain the modified Friedmann equation (\ref{eq:fried2})
\begin{equation}
    H^{2}(z) \simeq \frac{\rho_{\mathrm{m}}(z)}{6}\left[1+\sqrt{1+2\left( \frac{3\pi K}{\rho_{\mathrm{m}}(z)}\right)^{2}} \right].\label{eq:qcd1}
\end{equation}
Assuming the typical behavior for the CDM sector, $\rho_{\mathrm{m}}(z)=\rho_{\mathrm{m,0}}(1+z)^{3}$, in the limit $z\rightarrow -1$, we have
\begin{equation}
    H(z\rightarrow -1)\rightarrow \sqrt{\frac{\pi K}{\sqrt{2}}},\label{eq:qcd2}
\end{equation}
which is a constant value corresponding to de Sitter evolution. This latter result indicates that the model has a self-accelerated branch since $\rho_{\mathrm{m}}(z\rightarrow -1)\rightarrow 0$. It is worth mentioning this kind of cosmic evolution, as given by Eqs. (\ref{eq:qcd1}) and (\ref{eq:qcd2}), was also obtained within the context of a variable dark energy known as {\it ghost dark energy} in which the CDM contribution is considered \cite{ghost2, qcd}. 

From the result given above in Eq. (\ref{eq:qcd2}), we can establish a correspondence between the cosmological constant, $\Lambda$, and the Kaniadakis parameter by establishing $\Lambda/3 = \pi K/\sqrt{2}$. Then, in this case, we have
\begin{equation}
    \Omega_{\mathcal{K},0} = \frac{\sqrt{2}}{3\pi}\Omega_{\Lambda}. \label{eq:connection}
\end{equation}
As can be seen, the accelerated stage of cosmic evolution will be dominated by a de Sitter phase carried out by the parameter $K$ that mimics a cosmological constant. Notice that in this case, the well-known fundamental problem regarding the value of the cosmological constant can be solved since, in this case, $K$ emulates such a constant; therefore, $\Omega_{\mathcal{K},0}$ only should vary consistently with the observational range of $\Omega_{\mathrm{DE,0}}$. Therefore, in contrast to what was done in Ref. \cite{saridakis}, if we do not think in deviations from the $\Lambda$CDM model and treat the Kaniadakis cosmology as a new model, we observe that in the far future, the de Sitter phase can be accomplished from a quintessence type evolution in a universe filled with only CDM at late times. 

%%%%%%%%%%%%%%%%%%%%%%%%%%%%%%%%%%%%%%%%%%%%%%%%%%%%%%%%%%%
\subsection{Numerical estimation of the $K$-parameter with observational parameters}
\label{sec:kaniadakis3}
%%%%%%%%%%%%%%%%%%%%%%%%%%%%%%%%%%%%%%%%%%%%%%%%%%%%%%%%%%
In order to provide a simple connection between the theoretical framework and observational data, in the following we present a numerical exercise based on current cosmological parameters. Although a complete test against observations is beyond the scope of this work, this illustrative example demonstrates the viability of the Kaniadakis parameter $K$ in reproducing an accelerated expansion consistent with observations. For this purpose, we consider the following values for the parameters involved: Hubble constant $H_0= 70 \ \mbox{km} \ \mbox{s}^{-1} \mbox{Mpc}^{-1} \simeq 2.27 \times 10^{-18} \mbox{s}^{-1}$, $\Omega_{\mathrm{m,0}}=0.3$, and a deceleration parameter value consistent with $\Lambda CDM$ model, $q_0 =-0.5$ \cite{planck1518, data}. The entropy is simply given by $S_{\mathrm{BH,0}}=\frac{\pi}{H_{0}^{2}}$, as can be seen in Eq. (\ref{eq:entropy}). At first order $\mathcal{K}$ we can estimate $\mathcal{K}=0.82$ from equation (\ref{eq:decel}), therefore, from our definitions given above we obtain $\Lambda=  3\pi K /\sqrt{2}\simeq 9 \times 10^{-36} \  \mbox{s}^{-2}$, where we have considered $\mathcal{K}:=K\pi H^{-2}_{0}$, we can see that our value is in good agreement with the interval established by the Planck 2018 and Pantheon$+$ collaborations for $\Lambda$, which situates the cosmological constant between the values $9\times 10^{-36} \ \mbox{s}^{-2}$ and $12 \times 10^{-36} \ \mbox{s}^{-2}$, approximately \cite{planck1518, Brout:2022vxf}.

It is important to mention that in the inflationary context it was found that the $K$-parameter must obey the condition $K \lesssim \mathcal{O}(10^{-1})$ to be consistent with the BICEP and Planck data. This is not contradictory with the existing literature and the present work, but this result reveals a significant discrepancy when comparing different scenarios, for example, the value $K \simeq \mathcal{O}(10^{-83})$ was obtained from baryogenesis \cite{luciano1} and from BAO measurements it was estimated around $K \simeq \mathcal{O}(10^{-125})$ \cite{almada}. As stated in \cite{inflation}, this behavior suggests a generalized scenario where the entropy could be described by a dynamic $K$-parameter capable of providing a unified explanation for the different values of $K$ observed at distinct cosmic epochs. The concept of a variable parameter has previously been explored in the context of Tsallis entropy, a formalism strongly motivated by quantum field theory \cite{exponent}. This suggests that an interplay between quantum gravity and Kaniadakis entropy could address the aforementioned gap problem, which presents a promising avenue for future research.

In summary, the value for the cosmological constant derived from our framework is found to be of the same order of magnitude as current observational datasets estimate, we emphasize that no direct observational fitting has been performed. Hence, the current results should be viewed as preliminary. A full comparison—including angular diameter distance relations, the CMB acoustic scale, and type Ia supernovae luminosity distances—will be addressed in future work to better assess the observational viability of Kaniadakis-modified Friedmann dynamics. This outcome reinforces the potential of this thermodynamic approach to reproduce key features of the standard cosmological model. Further refinement through direct observational fitting would likely reduce the residual differences, strengthening model validation.

To end this section we stress that our findings align with previous studies exploring cosmological models based on Kaniadakis entropy. For instance, in Ref. \cite{npb1} the authors derived modified Friedmann equations based on the holographic principle for a non-flat universe, yielding a deformation term that effectively drives cosmic acceleration. Similarly, in Ref. \cite{npb2} the authors explored the usual and dual Kaniadakis entropies and confirmed their observational viability to solve some of the tensions already known in modern cosmology. Although our approach and derivation of the dynamics are distinct, the resulting modification to the Friedmann equations exhibits a key structural similarity: the emergence of late-time acceleration governed by the entropy deformation parameter, $K$. This feature reinforces the potential of the Kaniadakis formalism as a compelling explanation for dark energy. We emphasize that our framework is purely thermodynamic, in contrast to the aforementioned references. Overall, our results are qualitatively consistent with these prior analyses, lending further support to the hypothesis that generalized entropy formulations can explain the observed dark energy phenomenology.

%%%%%%%%%%%%%%%%%%%%%%%%%%%%%%%%%%%%%%%%%%%%%%%%%%%%%%%%%%%
\section{Conclusions}
\label{sec:final}
%%%%%%%%%%%%%%%%%%%%%%%%%%%%%%%%%%%%%%%%%%%%%%%%%%%%%%%%%%

In this work, we have investigated the cosmological implications of the Kaniadakis entropy within the framework of the gravity-thermodynamics conjecture. By applying this formalism to a spatially flat FLRW universe, we have discussed the modified Friedmann equations obtained from the deformed entropy-area relation and analyzed the resulting dynamics at the background level.

Our analysis reveals that the Kaniadakis entropy naturally leads to an effective dark energy component capable of explaining the observed late-time cosmic acceleration in a universe filled with a CDM fluid without invoking an explicit cosmological constant. The acceleration arises purely from the entropy-induced geometrical corrections, and in the far future, the model tends toward a de Sitter phase. Therefore, Kaniadakis cosmology is free of some issues inherent to the cosmological constant when interpreted as a fundamental constant of gravitational theory. In other words, the deformation parameter $K$ itself can emulate a cosmological constant, driving the late times acceleration. This is established by the explicit correspondence given in Eq. (\ref{eq:connection}), which shows that the observed value of $\Lambda$ can be reproduced by a purely entropic origin and that expression (\ref{eq:kanapprox}) is viable to describe the late times behavior of the universe. This represents a conceptual shift: rather than introducing both $\Lambda$ and $K$ as independent sources, we reinterpret the Kaniadakis parameter as the fundamental driver of cosmic acceleration. Such a reformulation sheds new light on the cosmological constant problem, pointing towards a statistical–thermodynamic origin for the accelerated expansion

Additionally, we showed that the second law of thermodynamics and the convexity condition are satisfied throughout the cosmic evolution, ensuring thermodynamic consistency. On the other hand, we also found that the model exhibits similarities to some other dynamical dark energy models with the following two important characteristics: the existence of a self-accelerated branch and de Sitter evolution recovered in the far future. This latter result is particularly interesting, since the identification of the Kaniadakis deformation parameter $K$ with an effective cosmological constant in the asymptotic future offers an alternative route to interpret the origin of cosmic acceleration, potentially alleviating the theoretical challenges associated with the fine-tuning of $\Lambda$ in standard cosmology. However, several important questions remain open for future investigation. As found in this work, the precise physical interpretation of the $K$-parameter in a cosmological context requires further exploration, as does the potential connection between this formalism and quantum gravity approaches.\\

In light of these results, we recognize several directions for future work:
\begin{itemize}
    \item \textbf{Perturbative analysis}: While our current investigation was focused on the homogeneous background dynamics, a complete cosmological model must also address linear perturbations. Generalized entropy formalisms may affect the growth of structures and CMB anisotropies. A detailed study of perturbative dynamics in Kaniadakis cosmology is therefore essential to assess its viability compared to the $\Lambda$CDM model and to break the degeneracy between the de Sitter phases.
    \item \textbf{Observational constraints}: Although we have highlighted the potential of $K$ to induce a self-accelerating universe and possibly ease observational tensions such as the Hubble constant discrepancy, no direct comparison with observational data has been made here. A statistical analysis using SNe Ia, BAO, and CMB data to constrain $K$ and assess the fit quality remains a task for future work.
    \item \textbf{Fundamental interpretation of $K$}: The physical origin of the Kaniadakis parameter remains to be understood at a deeper level. While it arises from a generalized relativistic statistical framework, its appearance in gravitational and cosmological contexts may point to more fundamental structures—possibly connected to quantum gravity, holographic scenarios, or spacetime microstructure, to mention some.
\end{itemize}
In conclusion, the Kaniadakis entropy provides a compelling and thermodynamically consistent mechanism for driving cosmic acceleration, opening promising alternatives for reinterpreting dark energy from an entropic perspective. We hope that this work motivates further exploration of generalized thermodynamic frameworks in cosmology, both at the theoretical and observational levels. This point will be crucial to fully validate the model.

\section*{Acknowledgments}
M.~Cruz work was partially supported by S.N.I.I. (SECIHTI-M\'exico). J.~Saavedra acknowledges the FONDECYT grant N°1220065, Chile. S.~Lepe acknowledges the FONDECYT grant N°1250969, Chile.

\appendix
\section{Series expansion of the entropy}
\label{sec:valuek}
The Bekenstein-Hawking entropy can be written in terms of the Hubble parameter as follows
\begin{equation}
    S_{\mathrm{BH}} = \frac{A}{4} = \frac{\pi}{H^{2}} = \frac{3\pi}{\rho}
\end{equation}
where the Friedmann constraint, $3H^{2}=\rho$, was used. If we consider a universe filled with matter and a cosmological constant, we can write the energy density, $\rho=\rho_{m,0}(1+z)^{3}+\Lambda$. Under these assumptions, we obtain the following limiting values for entropy:
\begin{align*}
& S_{\mathrm{BH}}\left( z\gg 0\right) \rightarrow \frac{3\pi }{\rho \left( z \gg 0\right) }, \\%
& S_{\mathrm{BH,0}} =\frac{3\pi }{\rho_{m,0}+\Lambda },\\ 
& S_{\mathrm{BH}}\left( z\rightarrow -1\right) \rightarrow \frac{3\pi }{\Lambda },
\end{align*}
which shows that $S_{\mathrm{BH}}\left( z\rightarrow -1\right)> S_{\mathrm{BH,0}} > S_{\mathrm{BH}}\left( z\gg 0\right)$; as the universe evolves from the past ($z\gg 0$) to the far future $z=-1$, the entropy increases. Using these results for $K \ll 1$, the expression (\ref{eq:kanapprox}) for the Kaniadakis entropy takes the form
\begin{equation}
S_{\mathrm{K}}(z) =\frac{3\pi }{\rho_{m,0}(1+z)^{3} +\Lambda }\left( 1+%
\frac{1}{6}\left( \frac{3\pi K }{\rho_{m,0}(1+z)^{3} +\Lambda }\right)
^{2}\right), 
\end{equation}
and remains valid under the condition $S_{\mathrm{BH}}>K^{2}S_{\mathrm{BH}}^{3}/6$, which in turn results as
\begin{equation}
   \frac{K^{2}}{6}\left[\frac{3\pi }{\rho_{m,0}(1+z)^{3} +\Lambda} \right]^{2} < 1, 
\end{equation}
if we specialize at the far future, we get
\begin{equation}
     \frac{K}{\Lambda} < \frac{\sqrt{6}}{3\pi} \simeq 0.26.
\end{equation}
This latter condition implies $\mathcal{O}(\Lambda) \gtrsim \mathcal{O}(K)$, which is sufficient to ensure the validity of the expression for the entropy given in (\ref{eq:kanapprox}). This is in good agreement with the findings discussed earlier in the work; see Eq. (\ref{eq:connection}). If we consider the present time case, we have
\begin{equation}
    \frac{K}{\rho_{\mathrm{m,0}}+\Lambda} < \frac{K}{\Lambda} < \frac{\sqrt{6}}{3\pi},
\end{equation}
therefore, despite the entropy increases as the universe evolves, the approximation given by Eq. (\ref{eq:kanapprox}) is valid for the description of the universe at late times.

\end{document}